\documentclass[leqno,11pt,twoside]{article}
\usepackage{epsfig}
\usepackage{graphicx}

\title{New developments of the methodology of the Modified method of simplest 
equation with application}
\author{Nikolay K. Vitanov}
\date{Institute of Mechanics, Bulgarian Academy of Sciences,
Akad. G. Bonchev Str., Bl. 4, 1113 Sofia, Bulgaria}

\begin{document}

\maketitle

\begin{abstract}
\noindent{\sc}
We discuss an extension of the modified method of simplest equation for 
obtaining exact analytical solutions of nonlinear partial differential 
equations.  The extension includes the  possibility for use of: (i) more 
than one simplest equation; (ii) relationship that contains as particular 
cases the relationship used by Hirota \cite{hirota} and the relationship 
used in the previous version of the methodology; (iii) transformation of 
the solution that contains as particular case the possibility of use of 
the Painleve expansion; (iv) more than one balance equation. The discussed 
version of the methodology allows: obtaining multi-soliton solutions of 
nonlinear partial differential equations if such solutions do exist and 
obtaining particular solutions of nonintegrable nonlinear partial differential 
equations. Examples for the application of the methodology are discussed.
\end{abstract}









 

\section{Introduction}
Differential equations occur in the process of mathematical study of many  
problems from natural and social sciences as these equations relate quantities 
to their changes and such relationships are frequently  encountered. 
Nonlinear differential equations are  used for modeling of processes in many branches 
of science such as fluid mechanics, atmospheric and ocean sciences, mathematical biology, 
social dynamics, etc. \cite{debn} - \cite{knvit}. In many cases the model equations 
are nonlinear partial differential equations and their exact solutions help us to 
understand complex nonlinear phenomena such as existence  and change of different 
regimes of functioning of complex systems, spatial localization, etc.  Because of the 
above the exact solutions of nonlinear partial differential equations are studied 
intensively \cite{ac} - \cite{tabor}. In the yearly years of the research on the 
methodology for obtaining exact solutions of nonlinear partial differential
equations one has searched for transformations that can transform the solved 
nonlinear partial differential equation to a linear differential equation. 
Numerous attempts for obtaining such transformations have been made and in 
1967 Gardner, Green, Kruskal and Miura \cite{gardner} managed to connect the 
Korteweg - de Vries equation to the inverse scattering problem for the linear 
Sch{\"o}dinger equation. This methodology is known today as \emph{Method of 
Inverse Scattering Transform}, \cite{ac}. Below we are interested in another line 
of research that was followed by Hirota who developed a direct method for obtaining 
such exact solutions - \emph{Hirota method} \cite{hirota}, \cite{hirota1}.
Hirota method is based on bilinearization of the solved nonlinear partial differential 
equation by means of appropriate transformations. Truncated Painleve expansions may lead
to many of these appropriate transformations \cite{tabor}, \cite{ct1}, \cite{wtk} and
the study of the applications of these truncated expansions \cite{k3}  
leaded to the formulation of the \emph{Method of Simplest Equation (MSE)} \cite{k05}, 
\cite{kl08}. We refer to the  articles of Kudryashov and co-authors for further 
results connected to \emph{MSE} \cite{k7} - \cite{k13}.
\par 
I have started my work on the method of simplest equation  by proposing the use of the 
ordinary differential equation of Bernoulli as simplest equation \cite{v10} and by 
application of the method to ecology and population dynamics \cite{vd10} where the concept 
of the balance equation has been used. Today the method of simplest equation has two versions.
The original version of Kudryashov is called \emph{Method of Simplest Equation - MSE} and there 
the determination of the truncation  of the corresponding series of solutions of the 
simplest equation is based on the first step in the algorithm for detection of the Painleve property.
An equivalent version is called \emph{Modified Method of Simplest Equation - MMSE} or \emph{Modified 
Simple Equation Method - MSEM} \cite{kl08}, \cite{vdk}, \cite{v11}. It if based on 
determination of the kind of the simplest equation and truncation of the series of 
solutions of the simplest equation by means of application of a balance equation. Up to 
now our contributions to the methodology and its application are connected to this version of 
the method \cite{v11a} - \cite{vdv17} and in \cite{vdv15} where we have extended the 
methodology of the \emph{MMSE} to simplest equations of the class
	\begin{equation}\label{sf}
		\left (\frac{d^k g}{d\xi^k} \right)^l = \sum \limits_{j=0}^{m} d_j g^j
	\end{equation}
where $k=1,\dots$, $l =1,\dots$, and $m$ and $d_j$ are parameters. The solution of Eq.(\ref{sf}) defines
a special function that contains as particular cases, e.g.:  trigonometric functions; hyperbolic functions;
elliptic functions of Jacobi; elliptic function of Weierstrass.
Our goal is to extend the methodology of \emph{MSE} and \emph{MMSE} in order to make it 
applicable to larger classes of nonlinear partial differential equations. 
\par

The text below is organized as follows.  In Sect. 2  we discuss a version the modified 
method of simplest equation that makes the methodology  capable to obtain multi-soliton 
solutions of nonlinear partial differential equations. Sect.3 contains examples of applications
of the method. Several concluding remarks are given in Sect. 4.
\section{Extended version of the modified method of simplest equation}
In the previous version of the method  we have used a representation of the searched 
solution of a nonlinear partial differential equation as power series of a solution of a 
simplest equation. This approach does not work for the case of search for bisoliton, 
trisoliton, and multisoliton solutions because the previous version of the modified 
method of simplest equation  was connected to the use of a single simplest equation. 
If we allow for use of more than one simplest equation then the modified method of 
simplest equation can be formulated in a way that makes obtaining of multisoliton 
solutions possible. Below we formulate such a version of the modified method of simplest equation. 
The schema of the old and the new version of the methodology is shown in Fig. 1.
\begin{figure}[!htb]
	\centering
	\includegraphics[scale=0.5]{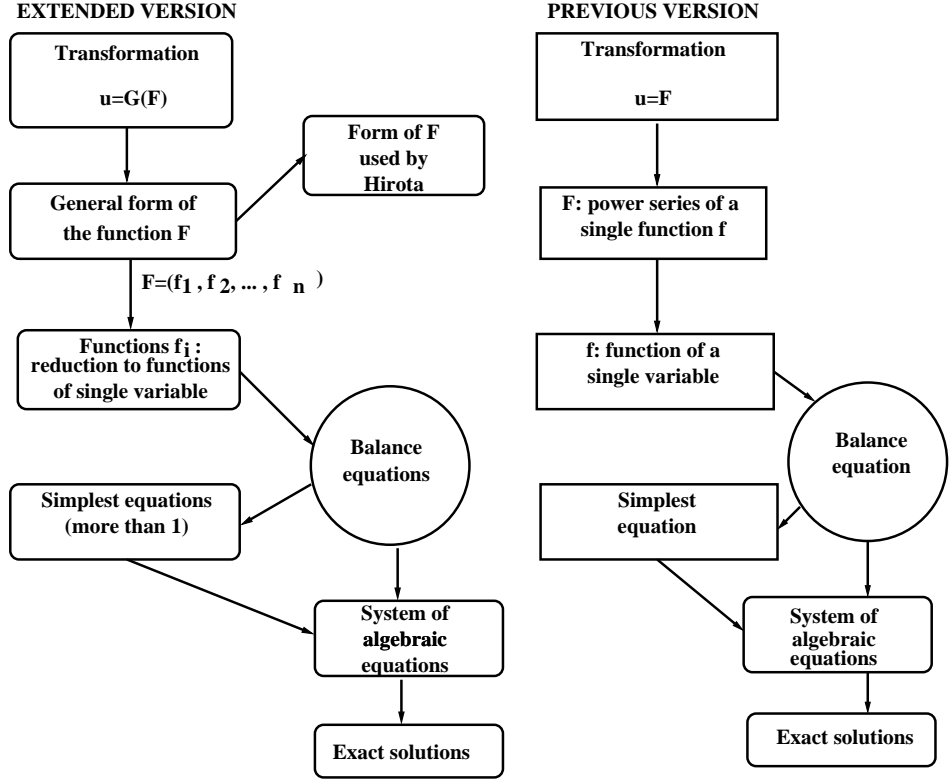}
	\caption{Schema of the previous version (righ-hand side) and of the extended
		version (left-hand side) of the Modified method of simplest equation. 
		The transformation used in the extended version of the method contains as 
		particular case the transformation used in the previous version of
		the method. The form of the function $F$ used in the extended version
		of the method contains as particular case the form of the function
		used in the previous version of the method. The new form of the function $F$ 
		contains as particular case also the form of the function used by Hirota. 
		In the extended form of the method more than one simplest equation can be used 
		and more than one balance equation can arise. In the previous version of the method 
		one uses one simplest equation and one balance equation.}
\end{figure}
\par
Let us consider a nonlinear partial differential equation 
\begin{equation}\label{eq}
	{\cal{R}}(u,\dots)=0
\end{equation}
where ${\cal{R}}(u,\dots)$ depends on the function $u(x,...,t)$
and some of its derivatives participate in  ($u$ can be a function of more than 1 spatial coordinate).
The 7 steps of the extended modified method of simplest equation are as follows.\\
\textbf{Step 1.)}
	We apply a transformation
	\begin{equation}\label{m1}
		u(x,\dots,t)=H(F(x,\dots,t))
	\end{equation}
	where $H(F)$ is some function of another function  $F$. In general
	$F(x,\dots,t)$ is a function of the spatial variables as well as of the time. The transformation $H(F)$
	may be the Painleve expansion \cite{hirota}, \cite{kudr90}, \cite{k3}, \cite{w1}, \cite{k10} or another 
	transformation, e.g., $u(x,t)=4 \tan^{-1}[F(x,t)]$ for the case of the 
	sine - Gordon equation, etc. \cite{mv1} - \cite{mv5}. In many particular cases one may skip this 
	step (then we have just $u(x,\dots,t)=H(x,\dots,t)$) 
	but in some cases the step is necessary
	for obtaining a solution of the studied nonlinear PDE. The application of Eq.(\ref{m1}) to 
	Eq.(\ref{eq}) leads to a nonlinear PDE for the function $F(x,\dots,t)$.\\
\textbf{Step 2.)}
	The function $F(x,\dots,t)$ is represented as a function of other functions $f_1,\dots,f_N$ ($N=1,2,\dots$)
	that are  connected to solutions of some differential equations (these equations can be partial 
	or ordinary differential equations) that are more simple than Eq.(\ref{eq}). We stress that
	the forms of the function $F(f_1,\dots,f_N)$ can be different. One example is
	\begin{eqnarray}\label{m2}
		F &=& \alpha + \sum \limits_{i_1=1}^N \beta_{i_1} f_{i_1} + \sum \limits_{i_1=1}^N  \sum \limits_{i_2=1}^N 
		\gamma_{i_1,i_2} f_{i_1} f_{i_2} + \dots + \nonumber \\
		&&\sum \limits_{i_1=1}^N \dots \sum \limits_{i_N=1}^N \sigma_{i_1,\dots,i_N} f_{i_1} \dots f_{i_N}
	\end{eqnarray}
	where $\alpha,\beta_{i_1}, \gamma_{i_1,i_2}, \sigma_{i_1,\dots,i_N}\dots  $ are parameters.
	The relationship (\ref{m2}) contains as particular case the 
	relationship used by Hirota \cite{hirota}. The power series $\sum \limits_{i=0}^N \mu_n f^n$ (where
	$\mu$ is a parameter) used in the previous versions of the methodology of the modified method of 
	simplest equation are a particular case of the relationship (\ref{m2}) too.\\
\textbf{Step 3.)} 
	In general the functions $f_1,\dots,f_N$ are solutions of partial differential equations.
	By means of appropriate ans{\"a}tze (e.g.,  traveling-wave ans{\"a}tze such as 
	$\xi = \hat{\alpha} x + \hat{\beta} t$; $\zeta =\hat{\gamma} x + \hat{\delta} t$, 
	$\eta = \hat{\mu} y + \hat{\nu}t \dots$) 
	the solved   differential equations for $f_1,\dots,f_N$ may be reduced to   differential equations 
	$D_l$, containing derivatives of one or several functions
	\begin{equation}\label{i1}
		D_l \left[ a(\xi), a_{\xi},a_{\xi \xi},\dots, b(\zeta), b_\zeta, b_{\zeta \zeta}, \dots \right] = 0; \ \
		l=1,\dots,N
	\end{equation}
	If the equations for the functions $f_1,\dots$ are ordinary differential equations one may skip this step 
	but the step may be necessary if the equations for $f_1,\dots$ are partial differential equations.\\
\textbf{Step 4.)}
	We assume that	the functions $a(\xi)$, $b(\zeta)$, etc.,  are  functions of 
	other functions, e.g., $v(\xi)$, $w(\zeta)$, etc., i.e.
	\begin{equation}\label{i1x}
		a(\xi) = A[v(\xi)]; \ \ b(\zeta) = B[w(\zeta)]; \dots
	\end{equation} 
We note  that the functions $A$ , $B$, $\dots$ are not prescribed. One may 
use a finite-series relationship, e.g., 
	\begin{equation}\label{i2}
		a(\xi) = \sum_{\mu_1=-\nu_1}^{\nu_2} q_{\mu_1} [v (\xi)]^{\mu_1}; \ \ \ 
		b(\zeta) = \sum_{\mu_2=-\nu_3}^{\nu_4} r_{\mu_2} [w (\zeta)]^{\mu_2}, \dots 
	\end{equation}
(where $q_{\mu_1}$, $r_{\mu_2}$, $\dots$ are coefficients) but other kinds of relationships may be used too. \\
\textbf{Step 5.)}
The functions  $v(\xi)$, $w(\zeta)$, $\dots$ 
are solutions of simpler ordinary differential equations called \emph{simplest equations}, i.e., 
the extended version of the methodology allows for the use of more than one simplest equation.\\
\textbf{Step 6.)}
	The application of the steps 1.) - 5.) to Eq.(\ref{eq}) transforms the left-hand side of 
	this equation. Let the result of this transformation  be a function that is a sum of terms where each 
	term contains some function multiplied by a coefficient. This coefficient contains some of the 
	parameters of the solved equation and some of the parameters of the solution. In the most cases
	a balance procedure must be applied in order to ensure that the above-mentioned relationships
	for the coefficients contain more than one term.
	This balance procedure may lead to one or more additional relationships \emph{balance equations} 
	among the parameters  of the solved equation and parameters of the solution. \\
\textbf{Step 7.)}
	We may obtain a nontrivial solution of Eq. (\ref{eq})  if all coefficients mentioned in Step 6.) are
	set to $0$. This condition usually leads to a system of nonlinear algebraic equations for the 
	coefficients of the solved nonlinear PDE and for the coefficients of the solution. Any nontrivial 
	solution of the above algebraic system leads to a solution the studied  nonlinear partial differential 
	equation. 
\section{Examples}
\subsection{The most simple example: bisoliton solution of the Korteweg-de Vries equation}
We shall describe this example very briefly in order to show the capacity of the extended
version of the methodology to lead to multi - soliton solutions of integrable partial
differential equations. We consider a version of the Korteweg - de Vries equation
\begin{equation}\label{a1}
u_t + \sigma u u_x + u_{xxx}=0
\end{equation}
where $\sigma$ is a parameter. The 7 steps of the application of the version of the modified 
method of simplest equation from Sect. 2 are as follows.\\
\textbf{ Step 1.)} \emph{The transformation}\\
We set $u=p_x$ in Eq.(\ref{a1}). The result is integrated  and we apply the transformation
$p=\frac{12}{\sigma} (\ln F)_x$. The result is
\begin{equation}\label{a2}
FF_{tx} + FF_{xxxx} - F_t F_x  + 3 F_{xx}^2 - 4 F_x F_{xxx} =0
\end{equation}
\textbf{ Step 2.)} \emph{Relationship among $F(x,t)$ and two functions $f_{1,2}$ that will be connected below to two simplest equations}\\ 
We shall use two functions
$f_1(x,t)$ and $f_2(x,t)$ and the relationship for $F$ is assumed to be a particular case of 
Eq.(\ref{m2}):
\begin{equation}\label{a3}
F(x,t) = 1 + f_1(x,t) + f_2(x,t) + c f_1(x,t) f_2(x,t) 
\end{equation}
where $c$ is a parameter. The substitution of Eq.(\ref{a3}) in Eq.(\ref{a2}) leads to
a nonlinear partial differential equation containing 64 terms.\\
\textbf{ Step 3.)} \emph{Equations for the functions $f_1(x,t)$ and $f_2(x,t)$}\\
The structure of the obtained allow us to assume  a the simple form of the equations for the functions $f_{1,2}$:
\begin{eqnarray}\label{a5}
\frac{\partial f_1}{\partial x} = \alpha_1 f_1; \ \ \ \frac{\partial f_1}{\partial t} = \beta_1 f_1; 
\frac{\partial f_2}{\partial x} = \alpha_2 f_2; \ \ \ \frac{\partial f_2}{\partial t} = \beta_2 f_2;
\end{eqnarray}
Eq.(\ref{a5}) transforms the solved nonlinear partial differential equation to a polynomial of $f_1$ and $f_2$. Further we assume that
$\xi = \alpha_1 x + \beta_1 t + \gamma_1$ and $\zeta = \alpha_2 x + \beta_2 t + \gamma_2$ and
$f_1(x,t) = a(\xi); \ \ \ f_2(x,t) = b(\zeta)$ where $\alpha_{1,2}$, $\beta_{1,2}$ and $\gamma_{1,2}$ are parameters.\\
\textbf{ Step 4.)} \emph{Relationships connecting  $a(\xi)$ and $b(\zeta)$ to the functions $v(\xi)$ and
$w(\zeta)$ that are solutions of the simplest equations} \\
In the discussed here case the relationships are quite simple. We can use Eq.(\ref{i2}) for the cases
$\mu_1 = \nu_2 = 1$ and $\mu_2 = \nu_4 = 1$. The result is:
$a(\xi) = q_1 v(\xi); \ \ \ b(\zeta) = r_1 w(\zeta)$.\\
\textbf{ Step 5.)} \emph{Simplest equations for $v(\xi)$ and $w(\zeta)$}\\
The simplest equations are
\begin{equation}\label{a8}
\frac{dv}{d\xi} = v; \ \ \ \frac{dw}{d\zeta} = w
\end{equation}
and the corresponding solutions are
$v(\xi) = \omega_1 \exp (\xi); \ \ \ w(\zeta)  = \omega_2 \exp(\zeta)$.
Below we shall omit the parameters $\omega_{1,2}$ as they can be included in the parameters
$q_1$ and $r_1$ respectively. We shall omit also $q_1$ and $r_1$ as they can be included in $\xi$ and $\zeta$.\\
\textbf{ Step 6.)} \emph{Transformation of the nonlinear PDE that contains 64 terms}\\
The substitution of all above in the nonlinear partial differential equation that contains 64 terms leads to 
a sum of exponential functions and each exponential function is multiplied by a coefficient. Each of these coefficients is a relationship
containing the parameters of the solution and all of the relationships contain more than one term. Thus
we don't need to perform a balance procedure.\\
\textbf{ Step 7.)} \emph{Obtaining and solving the system of algebraic equations}\\
The system of algebraic equations is obtained by setting of above-mentioned relationships to $0$.
Thus we obtain the following system:
\begin{eqnarray}\label{a10}
&& \alpha_1^3 + \beta_1 = 0, \ \ \ \alpha_2^3 + \beta_2 = 0, \nonumber \\
&& (c+1) \alpha_1^4 + 4 \alpha_2 (c-1) \alpha_1^3 + 6 \alpha_2^2 (c+1) \alpha_1^2 + [(4c-4)\alpha_2^3 + (\beta_1 + \beta_2) c + \nonumber \\
&& \beta_1 - \beta_2] \alpha_1 + [(c+1)\alpha_2^3 + (\beta_1 + \beta_2) c - \beta_1 + \beta_2] \alpha_2 = 0.
\end{eqnarray}
The non-trivial solution of this system is:  
$\beta_1 = -\alpha_1^3; \ \ \beta_2 = -\alpha_2^3; \ \
c = \frac{(\alpha_1 - \alpha_2)^2}{(\alpha_1 + \alpha_2)^2}$
and the corresponding solution of Eq.(\ref{a1}) is
\begin{eqnarray}\label{a12}
u(x,t) &=& \frac{12}{\sigma} \frac{\partial^2}{\partial x^2} \Bigg[ 1+ \exp \Big(\alpha_1 x - \alpha_1^3 t + \gamma_1 \Big) + \exp \Big(\alpha_2 x - \alpha_2^3 t + \gamma_2 \Big) + \nonumber \\
&& \frac{(\alpha_1 - \alpha_2)^2}{(\alpha_1 + \alpha_2)^2}
\exp \Big( (\alpha_1 + \alpha_2)x - (\alpha_1^3 + \alpha_2^3)t + \gamma_1 + \gamma_2 \Big) \Bigg]
\end{eqnarray}
Eq.(\ref{a12}) describes the bisoliton solution of the Korteweg - de Vries equation.
\subsection{ Second example: the generalized Maxwell-Cataneo equation}
As a second example we shall consider a nonlinear partial differential equation that is a generalization of the Maxwell-Cataneo kind of equation
\begin{equation}\label{ex2}
u_t+ru^au_{tt}=pu^{b-1}u_x^2+ q u^bu_{xx}
\end{equation}
The original Maxwell-Cataneo kind of equation is obtained when $a=0$, $b=1$ \cite{rever}.
We follow the methodology for the case $u(x,t) = H[F(x,t)] = F(x,t)$.
and search for traveling waves by using the traveling-wave ansatz 
$F(x,t) = F( \xi ) = F(\alpha x + \beta t)$ that reduces
the solved nonlinear PDE to a nonlinear ODE.
Then the solution $F(\xi)$ is searched as some function of another 
function $f(\xi)$, i.e.,  
\begin{equation}\label{i2}
F(\xi) = \sum_{\mu=-\nu}^{\nu_1} p_{\mu} [f (\xi)]^{\mu},
\end{equation}
$p_\mu$ are coefficients and $f(\xi)$ is a solution of
simpler ordinary differential equation (the  simplest equation):
\begin{equation}\label{se}
f_{\xi} = n \left[ f^{(n-1)/n} - f^{(n+1)/n} \right],
\end{equation}
where $n$ is \emph{an appropriate positive real number}.
The solution of this equation is $f( \xi ) = \tanh^{n}(\xi)$.
$n$ must be such real number that $\tanh^{n}(\xi)$ exists for $\xi \in (-\infty,
+ \infty)$ ($n=1/5$ is an appropriate value for $n$ and $n=1/4$ is not an
appropriate value for $n$). Following the steps of
the methodology we obtain two balance equations: $a = b = \frac{1}{n}$,
and the nonlinear partial differential equation (\ref{ex2})
is reduced to the system of nonlinear algebraic equations
\begin{eqnarray}\label{sys1}
r(n+1)\beta^2-[(p+q)n+q]\alpha^2 = 0 \nonumber \\
\Bigg \{r(n-1)\beta^2-\Bigg[ \Bigg((p+q)n-q \Bigg) \Bigg]\alpha^2 \Bigg \}\delta^{1/n}+\beta = 0 \nonumber \\
\Bigg[(p+q)\alpha^2-r \beta^2 \Bigg] n \delta^{1/n}-\frac{1}{2} \beta = 0.
\end{eqnarray}
One non-trivial solution of this system is
\begin{equation}\label{sl2}
\alpha=\frac{1}{2npr^{1/2}\delta^{1/n}}\Bigg[(n+1)(np+nq+q) \Bigg]^{1/2}; 
\ \  \beta = \frac{np+nq+q}{2npr\delta^{1/n}},
\end{equation}
and the corresponding solution of Eq.(\ref{ex2}) is
\begin{eqnarray}\label{knk}
u(x,t) = \delta \tanh^{n} \Bigg\{\frac{1}{2npr^{1/2}\delta^{1/n}}\Bigg[ \Bigg((n+1)(np+nq+q) \Bigg)^{1/2} x +
\frac{np+nq+q}{r^{1/2}}t \Bigg] \Bigg \}. \nonumber \\
\end{eqnarray}
Several particular cases are as follows. For $n=1$: the equation
\begin{equation}\label{ke1}
u_t+ruu_{tt}=pu_x^2+ q uu_{xx},
\end{equation}
has the solution
\begin{equation}\label{knk1}
u(x,t) = \delta \tanh \Bigg\{\frac{1}{2pr^{1/2}\delta}\Bigg[ \Bigg(2(p+2q) \Bigg)^{1/2} x +
\frac{p+2q}{r^{1/2}}t \Bigg] \Bigg \}.
\end{equation}
For $n=2$: the equation
\begin{equation}\label{ke2}
u_t+ru^{1/2}u_{tt}=pu^{-1/2}u_x^2+ q u^{1/2}u_{xx},
\end{equation}
has the solution
\begin{equation}\label{knk2}
u(x,t) = \delta \tanh^{2} \Bigg\{\frac{1}{4pr^{1/2}\delta^{1/2}}\Bigg[ \Bigg(3(2p+3q) \Bigg)^{1/2} x +
\frac{2p+3q}{r^{1/2}}t \Bigg] \Bigg \}.
\end{equation}
Finally let $n=1/3$. Then the equation
\begin{equation}\label{ke3}
u_t+ru^{3}u_{tt}=pu^{2}u_x^2+ q u^{3}u_{xx},
\end{equation}
has the solution
\begin{equation}\label{knk3}
u(x,t) = \delta \tanh^{1/3} \Bigg\{\frac{3}{2pr^{1/2}\delta^{3}}\Bigg[ 
\Bigg(\frac{4}{9}(p+4q) \Bigg)^{1/2} x +\frac{p+4q}{3r^{1/2}}t \Bigg] \Bigg \}.
\end{equation}
\section{Concluding remarks}
Above we  have discussed  an extended version of the methodology of the modified method
of simplest equation. The extension is  based on the possibility of use of more than 
one simplest equation,  on a transformation connected to  the searched solution and 
on a possibility of use of more general relationship among the solution of the solved 
nonlinear partial differential equation and the solutions of the simplest equations.
These possibilities  add the capability for obtaining multisolitons to the discussed 
extended methodology by keeping its ability to lead to particular exact solutions
of nonintegrable nonlinear partial differential equations.
Two examples of application of the methodology are 
presented and it is demonstrated that the balance procedure can lead to ore than one 
balance equation.

\end{document}